\documentclass[sigconf,natbib=true]{acmart}
\usepackage{xspace}
\AtBeginDocument{%
  \providecommand\BibTeX{{%
    \normalfont B\kern-0.5em{\scshape i\kern-0.25em b}\kern-0.8em\TeX}}}

\setcopyright{acmcopyright}
\copyrightyear{2018}
\acmYear{2018}
\acmDOI{10.1145/1122445.1122456}

\newcommand{\modelname}{Rank-LIME\xspace}

\acmConference[Woodstock '18]{Woodstock '18: ACM Symposium on Neural
  Gaze Detection}{June 03--05, 2018}{Woodstock, NY}
\acmBooktitle{Woodstock '18: ACM Symposium on Neural Gaze Detection,
  June 03--05, 2018, Woodstock, NY}
\acmPrice{15.00}
\acmISBN{978-1-4503-XXXX-X/18/06}

\usepackage{multirow}

\usepackage{natbib}
\begin{document}

\title{Rank-LIME: Local Model-Agnostic Feature Attribution \\ for Learning to Rank }


\author{Tanya Chowdhury}
\affiliation{%
  \institution{University of Massachusetts Amherst}
  \city{Massachusetts}
  \country{USA}
}
\author{Razieh Rahimi}
\affiliation{%
  \institution{University of Massachusetts Amherst}
  \city{Massachusetts}
  \country{USA}
}
\author{James Allan}
\affiliation{%
  \institution{University of Massachusetts Amherst}
  \city{Massachusetts}
  \country{USA}
}


\begin{abstract}
 Understanding why a model makes certain predictions is crucial when adapting it for real world decision making. LIME is a popular model-agnostic feature attribution method for the tasks of classification and regression. However, the task of learning to rank in information retrieval is more complex in comparison with either classification or regression. In this work, we extend LIME to propose Rank-LIME, a model-agnostic, local, post-hoc linear feature attribution method for the task of learning to rank that generates explanations for ranked lists. 
  We employ novel correlation-based perturbations, differentiable ranking loss functions and introduce new metrics to evaluate ranking based additive feature attribution models. We compare Rank-LIME with a variety of competing systems, with models trained on the MS MARCO datasets and observe that Rank-LIME outperforms existing explanation algorithms in terms of  Model Fidelity and Explain-NDCG.  With this we propose one of the first algorithms to generate additive feature attributions for explaining ranked lists.
 \end{abstract}


\maketitle

\section{Introduction}
A large number of explanation methods have been introduced  in the last few years~\cite{ribeiro2016should,lundberg2017unified,shrikumar2017learning}. While some are motivated by generating explanations of deep neural models for end users, others are built in order to gain an understanding of unknown properties and mechanisms of the underlying data-centric process. Explanations that belong to the class of additive feature attribution methods (e.g. LIME and SHAP) have played a large role in supporting the above goals, especially for tasks like regression and classification.

Explanation of learning-to-rank models, though, has received little attention. 
Singh and Anand.~\cite{singh2019exs} extended LIME to explain the relevance of a document to a query. However they (i)~do not evaluate the quality of generated explanations, and (ii)~do not explain the reason behind the order of documents retrieved for a query by the learning-to-rank model. Singh et al.~\cite{singh2020valid} proposed a greedy approach to identify the top features responsible for generating a ranked list. However, their approach (i)~has only been suggested for  rankers with human-engineered features as input, and 
(ii)~does not scale well for a large number of features.
In contrast, we see great value in a general, post-hoc ranking explanation techniques to understand the behaviour of deep learning-to-rank models. 

It might be that existing point-wise explainers can be used to explain a ranked list one document at a time. However, they focus on features that are responsible for a particular document being relevant at a time. In our early experiments, we discovered that those features could not be used to reconstruct the original ordering reliably. Instead, a list-wise approach has the potential to identify exactly which features are important to achieve a particular ordering of documents. We believe such an explanation is thus more useful to a user who wants to understand the rationale behind an ordering (e.g., a product search result).

We describe \modelname, an approach to generate model-agnostic local additive feature attributions for the task of learning to rank. Given a black-box ranker whose architecture is unknown, a query, a set of documents and explanation features, and a small part of the training data,   \modelname returns a set of weights $w_i$ for the most important features, measuring the \textit{relative contribution} of those features towards deciding the ranking. We focus on generating \modelname explanations for state-of-the-art transformer-based rankers (e.g. BERT and T5), but \modelname can be used to generate explanations for other ranking models as well. We propose metrics based on Kendall's Tau and NDCG to compute the accuracy of the generated explanations for learning-to-rank models and compare them with strong baselines. 
 
 Our main contributions are (i) extending LIME to explain ListWise relevance functions, (ii) introducing correlation-based instance perturbations, (iii) employing ranking reconstruction loss functions in LIME, and (iv) proposing measures to evaluate feature attributions in ranking. To the best of our knowledge this is one of the first works on explaining listwise relevance functions using feature attributions. The scripts to reproduce our experiments will be freely available online.

\section{Related work}
LIME~\cite{ribeiro2016should} is one of the first steps towards model explainability in machine learning literature. The authors propose a model-agnostic linear explanation method where they locally approximate a classifier with an interpretable model by perturbing inputs and then generating labels for the perturbed inputs. The output of their model is a bar graph representing contributions of supporting and opposing features. Later,  \citet{lundberg2017unified} propose KernelSHAP, a framework similar to LIME, which satisfies properties of the classical Shapely values.

Explanations for neural models in Information Retrieval is a relatively unexplored task. One of the first works in this direction was EXS, a local post-hoc  explainability technique by \citet{singh2019exs}, where they extended the general LIME-classifier explainability model to  pointwise rankers. They use it to answer three types of IR explainability questions: 1)~Why is a given document relevant to a given query? 2)~Why is document A ranked higher than document B?, and 3)~What is the query intent learnt by the ranking model?. They do so by perturbing the inputs in the locality of the document of interest and then generating binary relevance judgements for these perturbed queries. These perturbed instances are fed to the LIME explainability model and visualized as in the original. 

Next, \citet{fernando2019study} explore a model-introspective explainability method for Neural Ranking Models (NRMs). They use the DeepSHAP~\cite{lundberg2017unified} model which in turn extends DEEP LIFT~\cite{shrikumar2017learning} to learn an introspective explainable model for deep neural rankers. They compare with NRM explanations generated by \citet{singh2019exs} and find that explanations generated by LIME and DeepSHAP are significantly divergent. 

Recently, \citet{singh2020valid} extend IR explainability approaches to human engineered features, and propose two metrics: \textit{validity} and \textit{completeness}. They try to optimize these two metrics by greedily finding a subset of the input model features, capping the set size at $K$, such that there is a high correlation between the rankings produced by the selected features and the original blackbox model, i.e., high validity. At the same time they try to maximize completeness, which they quantify as the negative Kendall's Tau correlation between non-explanation features and the original ranking. 

\citet{verma2019lirme} propose a model-agnostic pointwise approach to compute \textit{explanation vectors}, which can in turn be studied to find positive or negative contributions of a term towards a ranking decision. \citet{sen2020curious} extend that work to give explanations in terms of three primal IR features: \textit{frequency of a term in a document}, \textit{frequency of a term in a collection},  and \textit{length of a document} as the weights of a linear function. Other work \cite{zhang2019sigir,zhang2020explainable,wang2018explainable} delves into generating explainable decisions for recommender systems.

\section{Rank-LIME}

Let $f$ denote a black-box learning-to-rank model that, given a query $q$ and a set of documents $D = \{d_1, \dots, d_N\}$, returns a ranking $R$ of documents in $D$, i.e., $f(q,D) = R$.  The ranking model $f$ can be uni-variate or multivariate~\cite{10.1145/3397271.3401104}, trained with a pointwise, pairwise, or listwise loss function~\cite{10.5555/2018740}. 
The dominant approach for learning-to-rank models is a uni-variate scoring function~\cite{10.5555/2018740,mitra2018an}  where the ranker scores each document $d_i \in D$ individually and then sorts the documents based on their scores. However, as our goal is explaining a black-box ranker, we do not make any assumption on the ranker and thus want to explain the obtained ranking $R$ from a black-box learning-to-rank model $f$. 

Here, we focus on local explanation of the behavior of a learning-to-rank model for a single ranked list with respect to a query as proposed in LIME and subsequent works~\cite{ribeiro2016should,lundberg2017unified}. Specifically, the goal is to explain the ranking $R_x$ obtained from $f(.)$ for the single instance ${x = (q, D)}$.
Explanation models usually work on interpretable (or simplified) inputs $x'$ that map to the original inputs through a mapping function
as follows.
\begin{equation}
x = h_x(x').   
\label{eq:mapping-fn}
\end{equation}
Documents here are represented as a bag of words. For a vocabulary of $N$ words, each document is represented by an $N$-dimensional vector where the $i^{th}$ element represents the frequency of the $i^{th}$ word in that document. 

Extending LIME for the explanation of learning-to-rank models, we sample instances around $x'$ to approximate the local decision boundary of $f$. The perturbed instances are denoted by $z'$. 
Feeding the perturbed instance $z'$ to ranker $f(h_x(z'))$, one obtains the ranking $R_{z'}$.

Following LIME, We define an explanation as a model $g \in \mathcal{G}$ where $\mathcal{G}$ is the class of linear models, such that $g(z') = w_g \cdot z'$.
The \modelname explanation is then obtained by minimizing the following objective function. 
\begin{equation}
    \xi = \arg \min_{g \in \mathcal{G}} \mathcal{L}(f, g, \pi_{x'}) + \Omega(g),
    \label{ranklime-exp-opt-obj}
\end{equation}
where $\Omega(g)$ represents the complexity of the explainable model, and $\pi_{x'}$ measures the locality of perturbed instances. 
In LIME~\cite{ribeiro2016should}, the loss function $\mathcal{L}(.)$ is defined as the mean squared error, which is not applicable to the output of learning-to-rank models as ranking of documents. We discuss the choice of loss,  complexity, and locality functions in the following subsections.

\subsection{Locality Function}
The local kernel $\pi_{x'}$ in Eq.~\eqref{ranklime-exp-opt-obj} defines the locality of the instance to be explained to weight the perturbed instances. Following LIME~\cite{ribeiro2016should}, we use an exponential kernel as:
\begin{equation}
    \pi_{x'} (z') = \exp(-\Delta(x',z')^2/\sigma^2),
\end{equation}
where $\Delta(.)$ denotes a distance function between the instance to be explained and a perturbed instance in the space of explanation features, i.e.,  $x' = (q^{x'}, D^{x'})$ and $z' = (q^{z'}, D^{z'})$. 
We instantiate the distance function as:
\begin{equation}
    \Delta(x',z') = \delta(q^{x'}, q^{z'}) + \sum_{d_i \in D} \delta (d_i^{x'}, d_i^{z'}) ,
\end{equation}
where 
the function $\delta$ measures the cosine distance between the vector representations of queries or documents based on the explanation features.

\subsection{Feature Perturbations}
The LIME algorithm is largely susceptible to assigning improper attributions when its explanation features are large in number and correlated to each other~\cite{zhou2021s}. These become the largest inhibitors in using LIME for explaining ranking decisions, where the documents are often strongly correlated to each other. To counter this, we compute $\Sigma_C$, the covariance matrix of the features from a part of the training dataset and incorporate them to generate perturbations in the Rank-LIME algorithm.  

Perturbations in the original LIME algorithm are random. Each feature is sampled independently from a normal distribution centred around the instance with $\mu=0$ and $\sigma=1$.  However in Rank-LIME, we carry out perturbations in the vicinity of an instance $x$, having $\mu=0$ and $\Sigma=\Sigma_C$. These include both single feature perturbations as well as group perturbations. This helps maintain the feature correlations from the training set in the perturbed instances and avoids noisy off-manifold or out-of-distribution perturbations~\cite{slack2020fooling}. This is especially important in the presence of large documents and correlated features.

\subsection{Loss Functions}
The \modelname explanation model needs a differentiable loss function quantifying the divergence between two ranked lists, specifically the original ranking $R_x$ and the ranking $R_z$ for a perturbed instance. 
The metrics generally used to compare or evaluate ranked lists are Kendall's Tau and \emph{normalized discounted cumulative gain} (NDCG). However, neither of these metrics are differentiable, and hence cannot be used to train a regression model to generate explanations.  We thus use a proxy of the above non-differentiable metrics in the \modelname framework as follows:

\noindent \textbf{ListNet}~\cite{cao2007learning} represents each ranked list with a probability distribution -- top-1 probability as an approximation of permutation probability. It then uses cross-entropy to measure the dissimilarity between two ranked lists.\\
\textbf{RankNet}~\cite{chen2009ranking} is a pairwise probabilistic loss function, aggregated to calculate list comparison scores.\\
\textbf{ApproxNDCG}~\cite{qin2010general} is a listwise proxy loss to NDCG that replaces the position and truncation functions of NDCG with a smooth function based on the position of documents. \\
\textbf{NeuralNDCG}~\cite{Pobrotyn2021NeuralNDCG} is a new proxy to NDCG that uses a differentiable approximate alternative to the sort function in NDCG, which is then plugged into the NDCG formula to compute relevance. Its neural sorting function is able to sort documents with high accuracy and bounded error rates.  \\

\section{Experiments}
\label{sec:exp}

\textbf{Datasets}
For experiments on text-based ranking models, we consider BM25, BERT~\cite{devlin-etal-2019-bert}, and T5~\cite{2020t5} ranking models as the black-box rankers to be explained. BERT and T5 rankers are fine-tuned using the MS MARCO  passage ranking dataset~\cite{nguyen2016ms} following previous studies~\cite{nogueira2019passage}. To conduct our experiments, we generate explanations for queries in the test set of the dataset. We generate explanations for the list of ten most relevant documents according to each ranker. We use the PyGaggle implementation of the BM25, BERT, and T5 rankers.\footnote{\url{https://github.com/castorini/pygaggle}}

\begin{table}[]
    \centering
    \caption{Comparing different perturbation methods and loss functions for Rank-LIME for word-based explanation of  the BERT-based ranker~\cite{nogueira2019passage}. Generation Time is measured in seconds.}
    \begin{tabular}{|c|c|c|c|}
    \hline
         Perturbation & Loss Function & Fidelity & Gen-Time \\ \hline
         \multirow{4}{*}{Single-Perturbations} & ListNet &  0.39 & \textbf{240} \\ \cline{2-4}
           & RankNet & 0.35 & 270\\ \cline{2-4}
          & ApproxNDCG  & 0.43 & 340 \\ \cline{2-4}
          & NeuralNDCG & 0.41 & 360\\ \hline
         \multirow{4}{*}{Group-Perturbations} & ListNet & 0.42 & 420\\ \cline{2-4}
           & RankNet & 0.39 & 440 \\ \cline{2-4}
          & ApproxNDCG  & \textbf{0.49} & 500 \\ \cline{2-4}
          & NeuralNDCG & 0.46 & 540\\ \hline

    \end{tabular}
    \label{tab:loss}
\end{table}

\noindent\textbf{Competing Methods.}
We compare the performance of the following explanation algorithms. We generate explanations by assigning linear attributions to the features chosen by each algorithm. 

\textbf{RANDOM} assigns random weights to $k$ of the explanation features, normalized to add up to 1.  

\textbf{Averaged-EXS} is proposed by~\citet{singh2019exs}. This approach addresses explainability for learning-to-rank models which use textual data as their input. They generate LIME attributions for each query-document pair based on various document relevance measures. We then aggregate their attributions of each document, select top $k$ features and normalize them to present as an attribution for the ranked list.


\textbf{Weighted-EXS}: One of the EXS methods uses LIME to  attribute why any \textit{document A is more relevant than document B}. We aggregate these explanations for each pair of documents and score them in a weighted manner, such that, the weight of each attribution is based on the difference in their document ranks in the original ranking. High rank difference  documents are given more weight than low rank difference ones. The top $k$ features from these weighted aggregated attributions are picked, which are then normalized such that they add up to 1.  

\textbf{TopKFeatures}, proposed by~\citet{singh2020valid}, introduces two metrics, validity and completeness, to select the top $k$ interaction features that contribute to decision making. This method however does not assign a weight to each feature, based on their contribution. As a result we cannot compare this method to our proposed method as is. Instead we assign a uniform weight value of $\frac{1}{k}$ to each feature in the result of this method, in order to enable comparison. \\

\noindent\textbf{Evaluation Metrics}
We propose the following metrics to evaluate the quality of explanation models. We derive intuition from similar metrics for the tasks of classification/regression and adapt them to suit ranking. 

   \textbf{Model Faithfulness/Fidelity} evaluates how good the explanation models are in reconstructing the black-box model's output for the given query.  We construct an \textit{explanation model's ranking} by linearly combining the features multiplied by their weights to form an ordering. We then compute the Kendall's Tau between the ranking by the black-box model and this obtained ordering.
    
    \textbf{Explain-NDCG@10} also evaluates how well the explanation model reconstructs the black-box model's output. Unlike Kendall's Tau, this metric is position sensitive and  yields higher scores for models which explain top ranked documents better. We use the scores obtained by the learning-to-rank model to assign NDCG relevance to each document. However, the scores of documents might be negative leading to an unbounded NDCG value. As a result, we use min-max normalization on the scores to decide on relevance labels while maintaining statistical significance, as recommended by Gienapp et al.~\cite{gienapp2020impact}. The higher the score, the more relevant the document. We then compute the NDCG of the reconstructed ranking with these relevance scores. We  average the value of NDCG@10 across all queries to report this metric. \\ 
    
  \begin{table}[t]
    \centering
    \caption{Comparing performance of different competing systems for word-based explanations of textual  rankers based on model fidelity and Explain-NDCG.}
    \begin{tabular}[width=\linewidth]{|c|c|c|c|}
    \hline
        \textbf{Ranker} & \textbf{System} & \textbf{Fidelity} & \textbf{Explain-NDCG}  \\ \hline 
        
        \multirow{5}{*}{BM25} & Random  & 0.20 & 0.0014  \\ \cline{2-4}
        & Average-EXS   & 0.45 & 0.2145 \\ \cline{2-4}
        & Weighted-EXS  & 0.56 & 0.3419 \\ \cline{2-4}
        & Top-K  & 0.39 & 0.1998  \\ \cline{2-4}
        & Rank-LIME  & \textbf{0.61} & \textbf{0.4019}  \\ \hline 
         
        \multirow{5}{*}{BERT} & Random  & 0.19 & 0.0002 \\ \cline{2-4}
        & Average-EXS  & 0. 39 & 0.1895 \\ \cline{2-4}
        & Weighted-EXS   & 0.42  & 0.2109 \\ \cline{2-4}
        & TopKFeatures  & 0.38 & 0.1034\\ \cline{2-4}
        & Rank-LIME  & \textbf{0.47} & \textbf{0.3210} \\ \hline 
         
        \multirow{5}{*}{T5} & Random  & 0.16 & 0.0042 \\ \cline{2-4}
        & Average-EXS & 0.32 & 0.1989  \\ \cline{2-4}
        & Weighted-EXS  & 0.41 & 0.2109  \\ \cline{2-4}
        & TopKFeatures  & 0.35 & 0.1487 \\ \cline{2-4}
        & Rank-LIME & \textbf{0.48} & \textbf{0.3415}  \\ \hline
    \end{tabular}
    \label{tab:token}
\end{table}

\noindent\textbf{Experimental Settings}
For explanations, we define the \textit{explanation feature set} as words of the set of documents $D$ and the query $q$ in the instance $x$ to be explained. We conduct experiments where ranking models BM25, BERT, and T5 are explained with features derived from the words of the instance to be explained. 
Perturbations are obtained by (i) modifying the frequency of a single feature at a time or
(ii) modifying the frequency of a group of features at a time.
The feature covariance matrix is computed using 1,000 randomly sampled documents from the training set. Each training document is represented as a  bag of words vector for $\Sigma_C$ to be computed.
We also compare different loss functions for different perturbation settings in Rank-LIME and report fidelity and relative explanation generation time for each scenario. 
We use the bag-of-words representation of  inputs to generate explanations. As a result the generated explanations are not position dependent. For all experiments, we choose 50 instances from the test sets of MS MARCO dataset to generate local explanations. We pick top $k$ ($k=8$) features for evaluating explanations by different systems in a fair manner.

\section{Results and Discussion}
\begin{figure}
    \centering
    \includegraphics[width=0.96\linewidth]{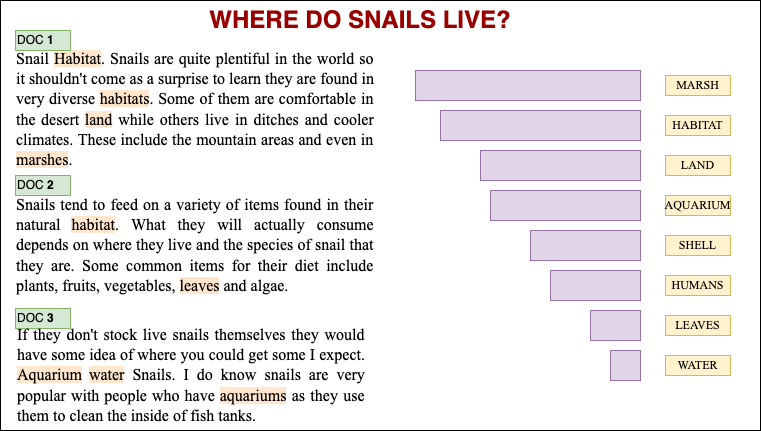}
    \caption{A toy example depicting Rank-LIME results on a query and three documents from the MS MARCO dataset. The documents were initially fed to the BERT Ranker and obtained the relevance scores, based on which they were ordered. Rank-LIME was later used to explain the model ordering and gives the bar chart shown to the right with its relative scores. Here we report the top 8 tokens which Rank-LIME found important in the model decision making.}
    \label{fig:example}
\end{figure}

\textbf{Rank-LIME Parameters.}
Table~\ref{tab:loss} shows the fidelity of \modelname in the word-based explanation of BERT-based ranker~\cite{nogueira2019passage} when different perturbations and loss functions are used. Single perturbations refer to perturbations where we perturb a single feature value at a time. Group perturbations refer to perturbations where we perturb groups of features in the instance being explained at a time.  We find the ApproxNDCG loss function outperforms other listwise loss functions proposed for generating Rank-LIME explanations by at least $4.8\%$. The group perturbation setting where we mask a subset of features at each perturbation achieves higher fidelity than the single perturbation scenario by $13.9\%$ but at the cost of generation time ($47\%$ overhead).  Since there is no time or budget constraint in this scenario, we use the ApproxNDCG loss function and group perturbation scenario for comparisons with later systems. Table~\ref{tab:loss} also depicts relative generation time for each kind of perturbation. We do not associate any absolute metric to it as generation time is implementation and system dependent.
Figure~\ref{fig:example} is an example Rank-LIME output corresponding to an instance of the MS MARCO dataset. 

\noindent \textbf{Rank-Lime vs Competing systems.}
Next, we compare Rank-LIME to competing systems in Table~\ref{tab:token}.  The reported results are computed over 50 instances randomly chosen from the test set of the dataset. We observe that Weighted-EXS outperforms Averaged-EXS by $24.4\%$. However, Rank-LIME outperforms the best baseline by $11.9\%$. We see an even bigger win for \modelname on Explain-NDCG, $17.6\%$ over the strongest baseline. A bigger win by Rank-LIME on Explain-NDCG as compared to Kendall's Tau (fidelity) intuitively suggests that \modelname pays more attention to explaining high scoring documents as compared to other explanation algorithms. We also observe that explanation algorithms achieve $29.7\%$ higher fidelity while explaining the BM25 ranker as compared to while explaining the BERT ranker. This suggests that LIME-based  algorithms that generate linear explanations perform worse  at explaining neural models with complex decision boundaries as compared to  explaining simple ranking models.

\section{Conclusion and Future Work}
The notion of explainability and how we quantify it is largely subjective. However, explanation models which are highly faithful as well as interpretable can be passed off as \textit{good} explanation models. In this work, we presented a  model-agnostic algorithm to provide linear feature attributions for results in the task of learning to rank. Our method is general and caters to pointwise, pairwise,  and listwise techniques. We proposed novel feature attribution techniques and evaluation metrics suitable for a ranking explanation task and showed that our method outperforms competing baselines.  Future work would incorporate causality in explanation generation in place of just considering correlation, since correlation does not always indicate causation. It would also be interesting to see how explanations from each of the competing systems impact human understanding of the blackbox model, via an user study.

\section{Acknowledgements}
This work was supported in part by the Center for Intelligent Information Retrieval, in part by NSF IIS-2039449 , and in part by NSF grant number 1813662. Any opinions, findings and conclusions or recommendations expressed in this material are those of the authors and do not necessarily reflect those of the sponsor.


\end{document}